\documentclass[12pt,twoside]{article}
\def\snr{SN\,1993J\,}
\def\sun{\odot}

\date{}
\usepackage {graphicx} 

\def\aap{A\&A\,  }
\def\apj{ApJ\,  }
\def\apjl{ApJ\,  }
\def\araa{ARA\&A  }

\def\mnras{MNRAS\,  }

\def\rmp{Rev. Mod. Phys.  }

\begin{document}
\centerline{\bf Adv. Studies Theor. Phys, Vol. x, 200x, no. xx,
xxx - xxx}

\centerline{}

\centerline{}
\centerline {\Large{\bf 
Direct conversion of the flux of kinetic energy 
 }}

\centerline{}

\centerline{\Large{\bf 
into radiation   in gamma-ray burst
}}

\centerline{}

\centerline{\bf {L. Zaninetti}}

\centerline{}

\centerline{Dipartimento  di Fisica Generale,}

\centerline{Universit\`a degli Studi di Torino,}

\centerline{via P. Giuria 1,  10125 Torino, Italy}

\begin{abstract}
The time evolution of a 
Gamma-ray burst  (GRB) is associated
with the evolution  of  
a  supernova remnant (SNR).
The time evolution of the flux of a GRB 
is  modeled introducing  a law 
for the density  of the  medium
in the advancing layer.  
The adopted  radiative   model  for the GRB  in the 
various  e.m.  frequencies 
is synchrotron emission.
The X-ray ring which appears a few hours 
after a GRB is simulated.
\end{abstract}


\section{Introduction}

Gamma-ray bursts (GRBs) started to be observed 
from an astronomical  point of view 
by \cite{Klebesadel1973}.
Recently some excellent reviews have been written,
see \cite{Paradijs2000, Meszaros2002,Piran2004,Gehrels2009}.
The recent trend of research connects 
GRBs with  supernovae (SN) explosions and their 
consequent 
remnants  (SNRs).
The afterglow is observed in X, optical, and infrared,
see \cite{Xin2010}.
The established connection between GRBs and SNRs   
allows of restricting 
the astrophysical environment to the advancing shock.
In the last  60 years the theoretical studies 
of the advancing shock of  SNR
have been focused on the
relationship  with the  instantaneous
radius of expansion, $R$,
which is of type  
$\propto~ t^m $, where $t$  is time
and $m$ is a parameter that depends on the chosen model.
The most popular model is  the Sedov--Taylor expansion 
which predicts  $R \propto  t^{0.4}$, 
see \cite{Sedov1946,Taylor1950a,Taylor1950b,Sedov1959},
and the thin layer approximation  in the 
presence of a constant density medium,
which predicts $R \propto  t^{0.25}$, 
see \cite{Dyson1997}.
A simple version of the Sedov--Taylor  solution, 
which also considers the transition from
the relativistic to the non-relativistic regime, appeared in
\cite{Huang1999,Paradijs2000}.
The radiative transfer 
during the afterglow phase
has  been analysed by 
\cite{Granot1999}, 
where a comprehensive
analysis, including images of the expected light curves, 
is carried out.
The actual situation of the theoretical and observational
research on SNRs
leaves a series of questions
unanswered or only partially answered:
\begin{itemize}
\item 
Can we deduce a theoretical expression for the 
counts/s versus time in a GRB assuming 
a direct proportionality  
between the flux of kinetic energy of the advancing shell
and the intensity of emission?
\item
Can we simulate  the X-ray ring
of GRB 032103,  
which  was observed  six hours after the burst?
\item 
Can we deduce a theoretical expression for the 
frequency at which the GRB becomes visible as 
a function of the time?
\end{itemize}
In order to answer  these questions, Section \ref{sec_data}
reports the data on GRBs.
Section  \ref{secsedov}
reviews 
the free expansion,
the Sedov--Taylor  solution 
and the power law model.
Section \ref{sec_transfer} reviews the existing situation 
about  the radiative transport equation
and a simple  geometrical  model  which 
leads to the ring emission.
Section \ref{sec_flux_time}
contains  some scaling arguments that lead 
to the observed  count/s versus time relationship.

\section{The Observations of GRBs}

\label{sec_data}

\begin{figure*}
\begin{center}
\includegraphics[width=10cm]{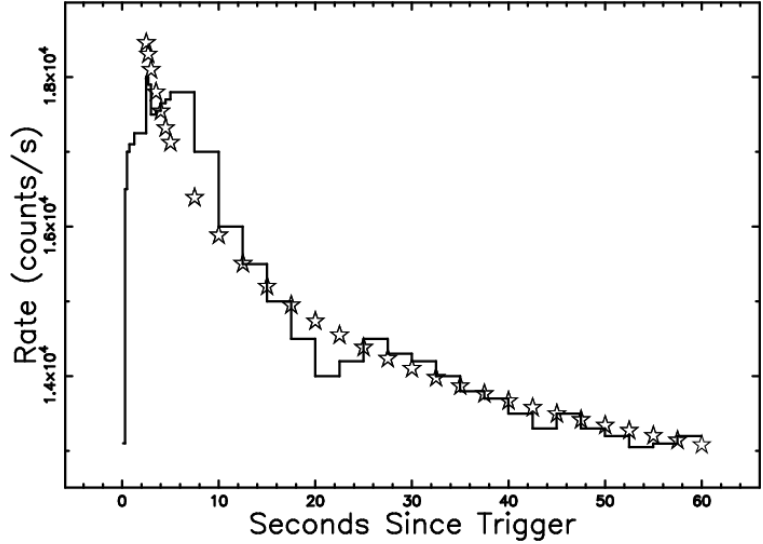}
\end {center}
\caption
{
Light curve of BATSE
trigger 8121 in the energy
channels 1--4 ( $>$ 20 kev)  (full line)
and power law  fit  (empty stars).
The real data have been extracted  by the author 
from the BATSE database available
at the web site 
http://www.batse.msfc.nasa.gov/batse/grb/lightcurve/~.
In this case $\delta =-0.1086$  and $C= 20382.27$
when  \quad $t~ \ge$ ~1.25~ s.
}
\label{kevtempo}
    \end{figure*}

GRBs are characterized by a rapid rise 
(called the trigger) in the rate of detected photons
per second  
at a given energy, for example 20 kev.
A typical  example 
observed by
the Burst and Transient Source Experiment (BATSE), 
see  \cite{Barthelmy2005},   on board 
the Compton Gamma Ray Observatory (CGRO)
 is reported in Fig.~\ref{kevtempo}.

>From the previous figure it is important to point 
out that the time necessary to reach the maximum
is  1.25~s.
In the following we will consider this time small
compared to the time in which  the counts/s  decreases.

A  numerical analysis of the observed  
rate  versus   time relationship 
can be done by assuming  a  power law 
dependence  for the rate of counts 
of the type
\begin{equation}
rate(t) = C t^{\delta}  ~counts~ s^{-1} 
\label{ratetime}
\quad ,
\end{equation}
where the two parameters $C$ and $\delta$ are 
found from a numerical analysis of the data.

\begin{figure*}
\begin{center}
\includegraphics[width=10cm]{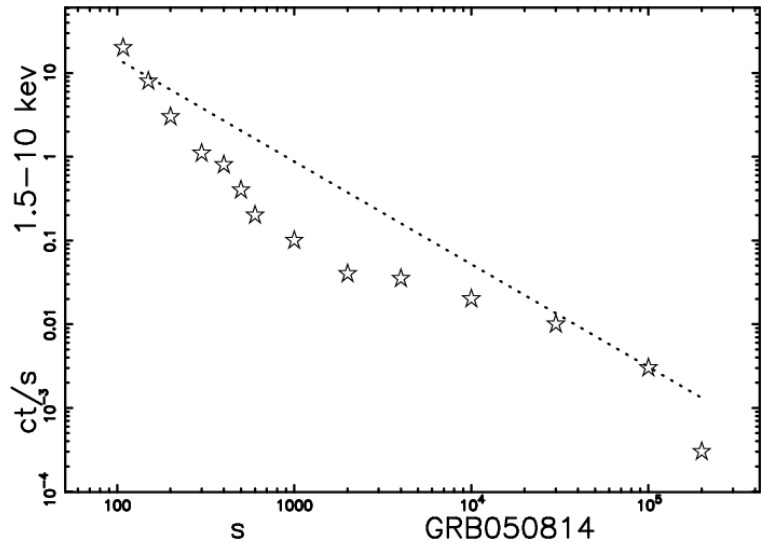}
\end {center}
\caption
{
Light curve of 
GRB050814 XRT 1.5--10 kev (empty stars)
and power law  fit       (dotted line).
The real data have been extracted  by the author 
from   
Fig. 3 (top-right)  
of Willingale et al. 2010.
In this case $\delta =-1.22$  and $C= 4214$.
}
\label{kevtempo2}
    \end{figure*}

An interesting analysis of the time evolution 
of 12 GRBs as observed by the
Swift Gamma-Ray Burst Mission      
(Swift), see \cite{Gehrels2004},
is reported in \cite{Willingale2010}.
We selected one GRB from the previous list,
namely,
GRB050814 XRT 1.5--10 kev 
which is visible in   Fig. 3 (top-right)  
of \cite{Willingale2010}.
For the previous  GRB as well the fit 
with relationship (\ref{ratetime}) is 
reported in Fig.~\ref{kevtempo2}.

\begin{figure*}
\begin{center}
\includegraphics[width=10cm]{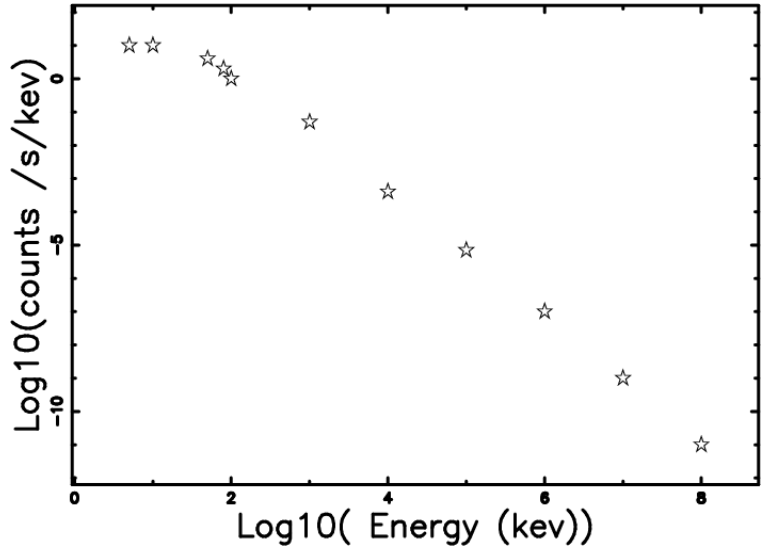}
\end {center}
\caption
{
Count  spectrum in energy  as  measured  
by the three 
instruments on board 
the Fermi Gamma-ray Space Telescope,
which are Nal, BGO, and LAT.
The data are extracted by the author from Fig. 2A 
in  \cite {Fermicoll2009}.
}
\label{spettro}
    \end{figure*}

The count spectrum   in energy  of GRB 080916C, see 
\cite {Fermicoll2009}, 
is  reported in Fig.~\ref{spettro}  and 
can be  approximated by the  following formula
\begin{equation}
E (kev) \approx 1563 E (kev) ^{-1.7}
\quad when   \quad  10~ kev~  \leq E \leq 10^7 kev. 
\end{equation}

\begin{figure*}
\begin{center}
\includegraphics[width=10cm]{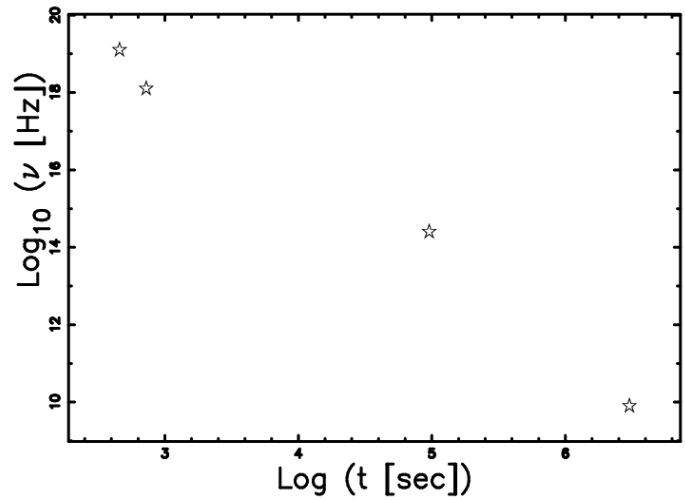}
\end {center}
\caption
{
Bands from the radio through the  IR/optical to X-ray
and BAT at  which  the expansion becomes optically 
thin.
The energy bands are expressed in logarithms of the frequency in Hz 
along the $y$-axis.
The  $x$-axis reports the logarithm of the time 
in seconds.
The data are extracted by the author from Table 1
in  \cite{Gou2007}. 
}
\label{freq_time}
    \end{figure*}

Another interesting property of the GRBs is that 
the frequency  of observation at which they 
start to be visible
decreases with time.
An example of such behaviour 
for  GRB 050904 in the BAT, XRT, J, and I bands
can be found in Table 1 
of  \cite{Gou2007}. The numerical analysis
of the data in Fig. \ref{freq_time} 
gives 
\begin{equation}
\nu = 1.03 10^{25} t(s)^{-2.27}\,  Hz
\label{trasparente}
\end{equation}
where $t$  has  been chosen 
as the value of time at which 
the flux at the chosen frequency  is  maximal.

Also interesting is  the case of  GRB 032103, 
which  was  observed in the X-ray
(energy range 0.7--2.5 keV) by
XMM-Newton's MOS cameras
six hours after the GRB, see \cite{Watson2004}.
In this case  
the  combined MOS images show 
two  concentric rings  which expand outwards.
The numerical  analysis gives
$\theta  \propto (t-t_0)^{1/2}$  
where $\theta$ is the angular separation between the
the two maxima in the radial profiles of counts,
$t$  is the time, and $t_0$ the initial time.
In the following we will associate the GRBs 
with the explosions of supernovae 
which are classified 
as SNR, see Section 4 in 
\cite{Piran2004}.

\section{Some existing solutions}

\label{secsedov}
This section  reviews the expansion  at 
constant velocity,
the Sedov--Taylor,  and the  power law models.

\subsection{The constant expansion velocity model}

The SNR  expands at a constant velocity until
the surrounding mass is
of the order of the solar mass.
This time, $t_M$, 
is 
\begin {equation}
t_M= 186.45\,{\frac {\sqrt [3]{{\it M_{\sun} }}}{\sqrt [3]{{\it n_0}}{\it 
v_{10000}}}} \quad \mathrm{yr}  
\quad ,
\end{equation}
where $ M_{\sun}$ is the number of solar masses 
in the volume occupied by the SNR,
$n_0$,  the
number density  expressed  in particles~$\mathrm{cm}^{-3}$,
and $v_{10000}$ the initial velocity expressed 
in units of 10 000 km/s, see \cite{Dalgarno1987}.

\subsection{The Sedov--Taylor solution}

\begin{figure*}
\begin{center}
\includegraphics[width=10cm]{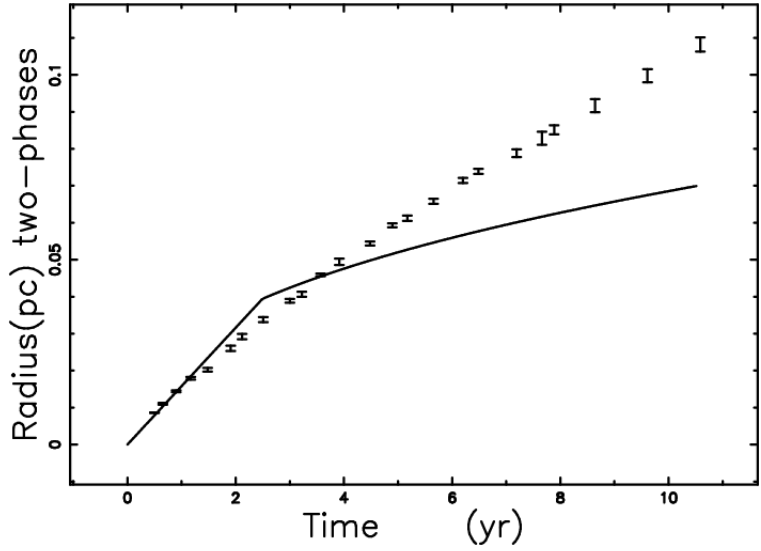}
\end {center}
\caption
{
Theoretical radius as given by the two-phase 
solution  
(full line),
and astronomical data of \snr with 
vertical error bars.
}
\label{1993duefasi}
    \end{figure*}

\ref{secsedov}
The  Sedov--Taylor  solution is     
\begin{equation}
R(t)=
\left ({\frac {25}{4}}\,{\frac {{\it E}\,{t}^{2}}{\pi \,\rho}} \right )^{1/5}
\quad , 
\label{sedov}
\end{equation}
where $E$ is the energy injected into the process
and $t$ is  time,
see~\cite{Taylor1950a,Taylor1950b,Sedov1959,Dalgarno1987}.
Our astrophysical  units are: time, ($t_1$), which
is expressed  in years;
$E_{51}$, the  energy in  $10^{51}$ \mbox{erg};
$n_0$,  the
number density  expressed  in particles~$\mathrm{cm}^{-3}$~
(density~$\rho=n_0$m, where m = 1.4$m_{\mathrm {H}}$).
In these units, Equation~(\ref{sedov}) becomes
\begin{equation}
R(t) \approx  0.313\,\sqrt [5]{{\frac {{\it E_{51}}\,{{\it t_1}}^{2}}{{\it n_0}}}
}~{pc}  
\quad .
\end{equation}
The Sedov--Taylor solution scales as $t^{0.4}$.
We are now ready to couple 
the Sedov--Taylor phase with the free expansion phase
\begin{equation}
 R(t)  = \left\{ \begin{array}{ll} 
0.0157\,{\it t}\, \mathrm{pc} & 
\mbox {if $t \leq 2.5$ yr } \\
0.0273\,\sqrt [5]{{{\it t}}^{2}} \, \mathrm{pc}   & 
\mbox {if $t >    2.5$\ yr.} 
            \end{array}
            \right.   
\label{twophases}
\end{equation}
This two-phase solution is obtained with
the following parameters 
$M_{\sun} = 1$, 
$n_0 =1.127 \,10^5$,  
$E_{51}=0.567$ 
and Fig. \ref{1993duefasi} reports its  temporal behaviour
as well as the data.
>From the previous figure, it is clear that the
free expansion plus the  Sedov--Taylor phase 
is not a satisfactory model.

\subsection{The power law model}

\begin{figure*}
\begin{center}
\includegraphics[width=10cm]{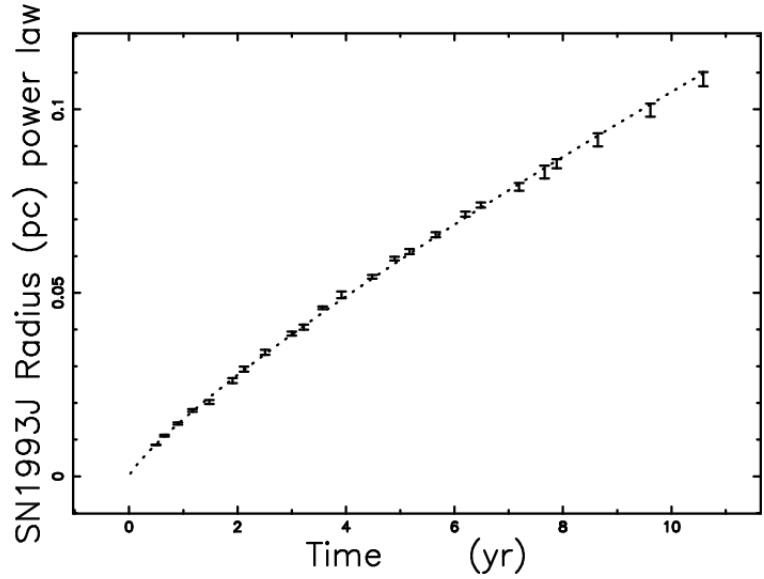}
\end {center}
\caption
{
Theoretical radius as given by the power law model
 (full line),
and astronomical data of \snr with
vertical error bars.
}
\label{radiust}
    \end{figure*}

We shall discuss a
power law
dependence  of the type
\begin{equation}
R(t) = C  t^{\alpha}
\label{rpower}
\quad ,
\end{equation}
where the two parameters 
$C$ and  $\alpha$ can be found
from the observations.
Ten years of observations  of  
\snr, see  \cite{Marcaide2009}, 
allow  of fixing these two parameters:
$\alpha = 0.828 \pm 0.0048$
and $C = 0.0155 \pm 0.00011$, 
and Figure  \ref{radiust}  displays  the data as well
as the fit.

\begin{figure*}
\begin{center}
\includegraphics[width=10cm]{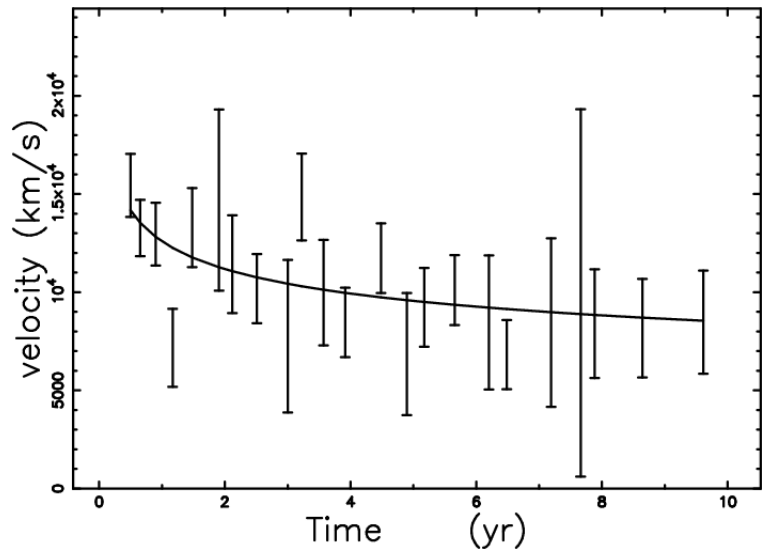}
\end {center}
\caption
{
Theoretical velocity
as  given  by Eq. (\ref{velpower})
(full  line)
and instantaneous velocity  of \snr  with
uncertainty.
}
\label{velt}
    \end{figure*}

This observed relationship
allows of expressing, from  an astrophysical point of 
view, the radius  as 
\begin{equation}
R(t) =   0.0155\,{t}^{ 0.828} \quad pc \quad,
\label{radiustime}
\end{equation}
where the  time  $t$  is  expressed in years.
The velocity  in this model  is  
\begin{equation}
V(t) = C \, \alpha t^{(\alpha-1)}
\label{vpower}
\end{equation}
and the astrophysical  version
\begin{equation}
V(t) = 12587.67\,{t}^{- 0.171} \frac{\mathrm{km}}{\mathrm{s}}  \quad .
\label{velpower}
\end{equation}
Figure \ref{velt} reports  the observed 
instantaneous 
velocity as deduced from the finite difference 
method and the best 
fit according to 
(\ref{velpower}).

\section{Radiative processes}
This section reviews the existing knowledge  
about
the transfer equation 
and a simple model 
which  produces an X-ring. 

\label{sec_transfer}

\subsection{The transfer equation}

The transfer equation in the presence of emission only,
see for example  
\cite{rybicki}
 or
\cite{Hjellming1988},
 is
 \begin{equation}
\frac {dI_{\nu}}{ds} =  -k_{\nu} \rho I_{\nu}  + j_{\nu} \rho
\label{equazionetrasfer}
\quad ,
\end {equation}
where  $I_{\nu}$ 
is the specific intensity, 
$s$ is the
line of sight, 
$j_{\nu}$ is the emission coefficient,
$k_{\nu}$ is a mass absorption coefficient,
$\rho$ is the density of mass at position $s$,
and the index $\nu$ denotes the frequency of
emission of interest.
The intensity of radiation, i.e., the photon flux,
is here identified with the counts at a given energy.
The solution to Equation~(\ref{equazionetrasfer})
 is
\begin{equation}
 I_{\nu} (\tau_{\nu}) =
\frac {j_{\nu}}{k_{\nu}} ( 1 - e ^{-\tau_{\nu}(s)} )
\quad  ,
\label{eqn_transfer}
\end {equation}
where $\tau_{\nu}$ is the 
optical depth at frequency $\nu$:
\begin{equation}
d \tau_{\nu} = k_{\nu} \rho ds
\quad.
\end {equation}
The framework of synchrotron emission,
as described in  sec. 4 of \cite{Schlickeiser},
 is 
often used in order to explain a GRB,
see for example 
\cite{Peer2006,Dermer2008,GRB090510_2010,Piran2010}.
The volume emissivity 
(power 
per unit frequency interval 
per unit volume per unit solid angle) 
of the ultrarelativistic radiation 
from a group of electrons, 
according to 
\cite{lang},  
 is  
\begin {equation}
\epsilon (\nu) =
\int  P(\nu) N(E) dE
\quad ,
\end{equation}
where $P(\nu)$ is the total power radiated 
per unit frequency interval by one electron
and  $N(E) dE$ is the number of electrons 
per unit  volume,  
per unit solid angle along the line
of sight, which are moving in the direction 
of the observer and whose energies lie in the range
$E$ to $E+dE$.
In the case of a power law  spectrum,
\begin{equation}
N(E)dE = K E^{-\gamma} dE  
\label{spectrum}
\quad  ,
\end{equation}
where $K$ is a constant.
The value  of the constant  $K$ can be found 
by assuming that  
the  probability density function 
(PDF, in the following)
for relativistic energy  
is of  Pareto type   
as defined in   \cite{evans}:
\begin {equation}
f(x;a,c) = {c a^c}{x^{-(c+1)}} \quad ,
\label{pareto}
\end {equation}
with $ c~> 0$.
In our case, 
$c=\gamma-1$ and  $a=E_{min}$,
where $E_{min}$ is the minimum energy.
>From the previous formula, we can extract
\begin{equation}
K= 
N_0 (\gamma -1) E_{min}^{\gamma -1} 
\quad  ,
\end{equation}
where $N_0$ is the total number
of relativistic electrons  per unit volume, 
here assumed to be approximately 
equal to the matter number density.
The previous  formula can also be expressed 
as  
\begin{equation}
K= 
\frac {\rho}{1.4 \, m_{\mathrm {H}}}
 (\gamma -1) E_{min}^{\gamma -1} 
\quad  ,
\end{equation}
where $m_H$ is the mass of hydrogen.
The emissivity  of  the ultrarelativistic  
synchrotron radiation from a homogeneous  
and isotropic distribution of electrons 
whose  $N(E)$  is given by 
Equation~(\ref{spectrum})
is, according to 
\cite{lang},  
\begin{eqnarray}
j_{\nu} \rho  =  \\
\approx 0.933 \times 10^{-23}
\alpha_L (\gamma) K H_{\perp} ^{(\gamma +1)/2 }
\bigl (
 \frac{6.26 \times 10^{18} }{\nu}
\bigr )^{(\gamma -1)/2 } \nonumber    \\
\mathrm{erg sec}^{-1} \mathrm{cm}^{-3} \mathrm{Hz}^{-1} \mathrm{rad}^{-2}
\nonumber   
\quad , 
\end{eqnarray}
where $\nu$ is the frequency
and   $\alpha_L (\gamma)$  is a slowly
varying function 
of $\gamma$ which is of the order of unity 
and is given by
\begin{equation}
\alpha_L(\gamma) =
2^{(\gamma -3)/2} \frac{\gamma+7/3}{\gamma +1}
\Gamma \bigl ( \frac {3\gamma -1 }{12} \bigr )  
\Gamma \bigl ( \frac {3\gamma +7 }{12} \bigr )  
\quad ,
\end{equation} 
for  $\gamma \ge \frac{1}{2}$. 

We now continue to analyse the case 
of an optically thin layer
in which $\tau_{\nu}$ is very small
(or $k_{\nu}$ is very small)
and where the density $\rho$ is replaced
by  the concentration $C(s)$
 of relativistic electrons:
\begin{equation}
j_{\nu} \rho =K_e  C(s)
\quad  ,
\end{equation}
where $K_e$ is a constant function
of the energy power law index,
magnetic field,
and frequency of e.m.  emission.
The intensity is now
\begin{equation}
 I_{\nu} (s) = K_e
\int_{s_0}^s   C (s\prime) ds\prime \quad  \mbox {optically thin layer}.
\label{transport}
\end {equation}
The increase in brightness
is proportional to the concentration 
integrated along
the line of  sight.
\subsection{The X-ring }

\label{ringsec}
\begin{figure*}
\begin{center}
\includegraphics[width=10cm]{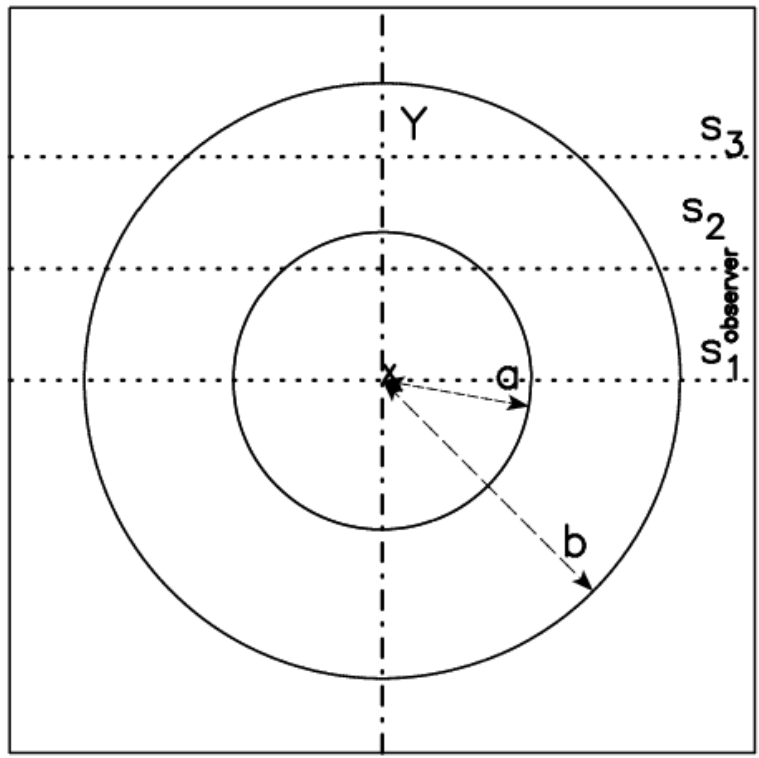}
\end {center}
\caption
{
The two circles (sections of spheres)  which   include the region
with constant density
are   represented through
 a full line.
The observer is situated along the $x$ direction, and 
three lines of sight are indicated.
}
\label{plotab}
    \end{figure*}

\begin{figure*}
\begin{center}
\includegraphics[width=10cm]{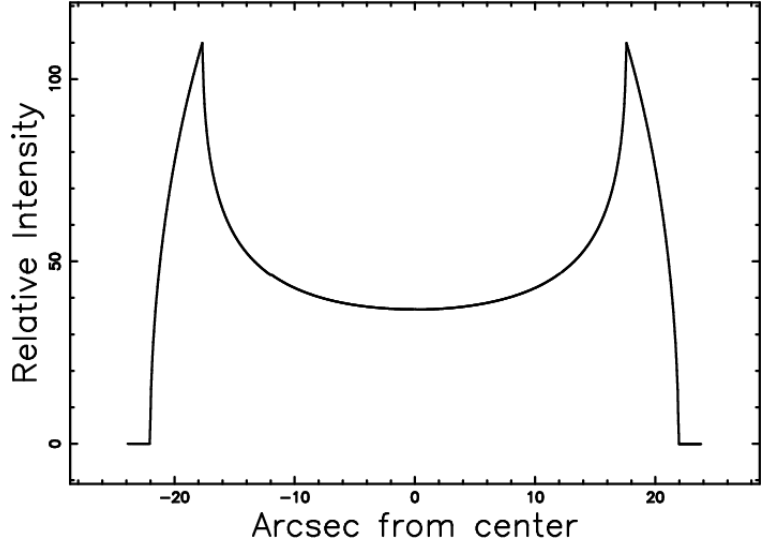}
\end {center}
\caption
{
 Cut of the mathematical  intensity ${\it I}$
 of the ring  model, Equation~(\ref{irim}), 
 crossing the centre    (full  line) of  GRB 031203 XMM-Newton 
 observation. 
 The $x$- and  $y$-axes  are in arcsec,
 $a=17.6  $ arcsec,  $b=22 $ arcsec
 and $\frac {I(y=a)} {I(y=0)}=3$
 as in Fig. 2 panel 6 of 
 \cite{Watson2004}.
}
\label{ring_cut}
    \end{figure*}

\begin{figure*}
\begin{center}
\includegraphics[width=10cm]{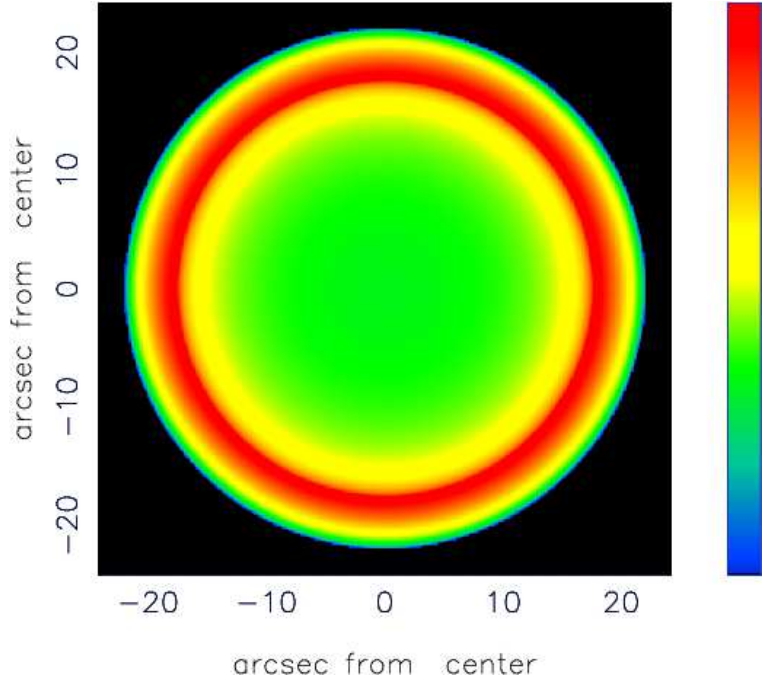}
\end {center}
\caption
{
Contour map  of  ${\it I}$
adjusted to simulate    
GRB 031203 XMM-Newton 
observation.
The $x$ and  $y$  axes  are in arcsec,
other parameters as in Fig.~\ref{ring_cut}.
}
\label{sfera}
    \end{figure*}

We assume that the number density 
of ultrarelativistic electrons 
$C$ is constant and in particular
rises from 0 at $r=a$ to a maximum value $C_m$, remains
constant up to $r=b$, and then falls again to 0.
This geometrical  description is reported in  
Fig.~\ref{plotab}.
The length of sight, when the observer is situated
at the infinity of the $x$-axis, 
is the locus    
parallel to the $x$-axis which  crosses  the position $y$ in a 
Cartesian $x--y$ plane and terminates at the external circle
of radius $b$.
The locus length is   
\begin{eqnarray}
l_{0a} = 2 \times ( \sqrt { b^2 -y^2} - \sqrt {a^2 -y^2}) 
\quad  ;   0 \leq y < a  \nonumber  \\
l_{ab} = 2 \times ( \sqrt { b^2 -y^2})  
 \quad  ;  a \leq y < b    \quad . 
\label{length}
\end{eqnarray}
When the number density
of ultrarelativistic electrons 
 $C_m$ is constant between two spheres
of radius $a$ and $b$ 
the intensity of radiation is 
\begin{eqnarray}
I_{0a} =C_m \times 2 \times ( \sqrt { b^2 -y^2} - \sqrt {a^2 -y^2}) 
\quad  ;   0 \leq y < a  \nonumber  \\
I_{ab} =C_m \times  2 \times ( \sqrt { b^2 -y^2})  
 \quad  ;  a \leq y < b    \quad . 
\label{irim}
\end{eqnarray}
The ratio between the theoretical intensity at the maximum $(y=a)$
 and at the minimum ($y=0$)
is given by 
\begin{equation}
\frac {I(y=a)} {I(y=0)} = \frac {\sqrt {b^2 -a^2}} {b-a}
\quad .
\label{ratioteorrim}
\end{equation}
The  parameter $b$ is identified with the external radius which denotes 
the transition from the perturbed medium.
The parameter $a$ can be found from 
the following formula:
\begin{equation}
a  = \frac
{
b \left(   (\frac {I(y=a)} {I(y=0)})_{obs}^2 - 1  \right) 
}
{
\left(   (\frac {I(y=a)} {I(y=0)})_{obs}^2 + 1  \right) }
\quad ,
\end{equation}
where  $(\frac {I(y=a)} {I(y=0)})_{obs} $ 
is the observed ratio between 
the maximum intensity at  the rim 
and the intensity at the centre.
A cut in the theoretical intensity 
of  GRB 031203 XMM-Newton
is reported in Fig.~\ref{ring_cut}
and  a  theoretical   image in Fig. \ref{sfera}.  

A comparison should be made with Fig.~7  in  \cite{Granot1999}.

\section{Flux density versus time }

\label{sec_flux_time}
The source of synchrotron luminosity 
is assumed here to be the flux of kinetic energy, 
$L_m$:
\begin{equation}
L_m = \frac{1}{2}\bar {\rho_L}  \pi R^2  V^3 
\quad ,
\end{equation}
where $R$  is the instantaneous radius of the SNR,
and $\bar {\rho_L}  $  is the averaged
 density in the advancing layer.
The   averaged density  
in the advancing layer 
can be parametrized as 
\begin{equation}
\bar {\rho_L}   =  
\frac  
{(\rho  \frac{4}{3} \pi R^3 )^{1/p} }
{ 4 \pi R^2 \Delta R  } 
\quad ,
\label{formularho}
\end{equation}
where $\Delta R$ is the thickness of the advancing radius
in which the mass resides.
In order to quantify the thickness 
we assume that 
\begin{equation}
\Delta R =  \frac{R(t)^{\eta_I}} {n} 
\quad  ,
\label{deltar}
\end{equation} 
with $\eta_I$   derived from 
the fit with the 
observations  and  $n$ 
a sufficiently large number 
such that 
\begin{equation} 
\frac{R(t_{max})^{\eta_I}} {n}  \ll R(t_{max}) 
\quad  ,
\end{equation}
where $t_{max}$ is the maximum time considered.

Therefore, the mechanical luminosity 
transforms to
\begin{equation}
L_m  \propto \bar{\rho_L} R^3 V^3
\quad .
\end{equation}
The temporal and velocity evolutions can be
given by the  power law 
dependencies of Equations (\ref{rpower}) and 
(\ref{vpower})
and therefore
\begin{equation}
L_m  \propto  t^{-\frac{6+{\eta_I} p}{3+p}}
\quad  .
\end{equation}
The  synchrotron luminosity
 $L_{\lambda}$  
and  the observed  flux  $S_{\lambda}$  
at a given wavelength       $\lambda$
are assumed   to be proportional  
to the  mechanical  luminosity,
and therefore
\begin{equation} 
S_{\lambda} =  S_0 (\frac {t}{t_{S0}})^{-\frac{6+{\eta_I} p}{3+p}}  
\quad ,
\label{fluxtime}
\end{equation}
where $S_0$ is the observed flux at  $t=t_{S0}$.
As an  example from the 
astronomical data  
of Fig.~\ref{kevtempo2},
we deduce that ${\eta_I}=1.095$.
The  progressive transparency
of the GRB to  the low  frequencies   with 
time 
can be explained by 
the dispersion relation of an e.m. 
wave in a cold plasma, which is 
\begin{equation}
\omega ^2 = \omega_{pe}^2 + k^2 c^2  
\quad ,
\label{dispersion}
\end{equation}
where 
$\omega$ is      the  frequency  of the e.m. wave,
$\omega_{pe}$ is the plasma frequency  of  the electrons,
$k$ is the wave-vector of the e.m. wave, and 
$c$ is the  velocity of light, see \cite{Schmidt1979}.
The plasma frequency of the electrons  is
\begin{equation}
\omega_{pe}=(4\pi n_ee^2/m_e)^{1/2}
\quad ,
\end{equation}
or
\begin{equation}
f_{pe}=\omega_{pe}/2\pi=8.98\times10^3{n_e}^{1/2}\,\rm Hz
\quad ,
\end{equation}
where $f_{pe}$ is the electron plasma frequency expressed in Hz.
A visual inspection of the dispersion relation
(\ref{dispersion}) 
allows us to say that in an
astrophysical environment in which the 
number density of the electrons is decreasing with time,
the frequency at which the medium becomes transparent 
will also progressively decrease.
The time dependence  of the  frequency  at  which the
medium is transparent, according to (\ref{formularho}),
is 
\begin{equation}
\nu  \propto t^
{
-1/2\,{\frac {-3+2\,p+{ \eta_{II}}\,p}{3+p}}
}
\quad ,
\end{equation}
where $\eta_{II}$ is the same  parameter
as in (\ref{deltar}) and the  index $II$ 
means a second way to deduce the free parameter.
A comparison of  the previous  formula with the observational
formula (\ref{trasparente})  
gives  $\eta_{II}=3.42$.

\section{Conclusions}

The  first two parts  of the law of motion
for a SNR are  thought to be a free expansion 
in which $R \propto  t$ and an energy  conserving  
phase, the so called Sedov--Taylor  phase,
in which   $R \propto  t^{2/5}$.
A careful analysis of \snr  in the first ten years
suggests instead that $R \propto  t^{0.82}$.

The scaling laws of the flux  of GRBs  with 
time can be consequently  deduced assuming 
that the  intensity of emission is  proportional
to the flux of the kinetic energy of the advancing 
shell, see Equation~(\ref{fluxtime}).
This result is reached assuming that 
the thickness of the advancing layer
increases  $\propto~R^{\eta}$.
The comparison between the observed flux  and the time  
of  GRB050814 XRT 1.5--10 kev    gives 
${\eta_I}=1.095$  and  the 
comparison between  the e.m. frequency of transparency  
and time 
gives   $\eta_{II}=3.42$.
The X-ray ring connected with GRB 031203 XMM-Newton
is simulated introducing two hypotheses:
(i) the emission of radiation is localized
    in a thin layer of the advancing expansion,
(ii) the emitting medium is supposed to be
     optically  thin. 


\end{document}